\documentclass[apjl]{emulateapj}
\usepackage{amsmath}
\usepackage{txfonts}
\usepackage{natbib}
\usepackage{graphicx}
\usepackage{color}

\begin{document}

\title{Erratum: ``Progenitor-explosion connection and remnant birth masses for 
neutrino-driven supernovae of iron-core progenitors'' (2012, ApJ, 757, 69)}

\shortauthors{Ugliano et al.}

% authors
\author{Thomas Ertl \altaffilmark{1,2}\\
  and\\
  Marcella Ugliano \altaffilmark{1},
  Hans-Thomas Janka \altaffilmark{1},
  Andreas Marek \altaffilmark{1}, and
  Almudena Arcones \altaffilmark{3,4}
  }

% affiliations
\altaffiltext{1}{Max-Planck-Institut f\"ur Astrophysik,
       Karl-Schwarzschild-Str. 1, 85748 Garching, Germany}

\altaffiltext{2}{Physik Department, Technische Universit\"at M\"unchen,
James-Franck-Stra\ss e 1, 85748 Garching, Germany}

\altaffiltext{3}{Institut f\"ur Kernphysik,
Technische Universit\"at Darmstadt,
Schlossgartenstr.~2,
64289 Darmstadt,
Germany}

\altaffiltext{4}{GSI Helmholtzzentrum
f\"ur Schwerionenforschung GmbH,
Planckstr.~1,
64291 Darmstadt,
Germany}

\maketitle
%\section{ }

%-----------------------------------------------------------------------
\begin{figure*}
\begin{center}
\includegraphics[width=\textwidth]{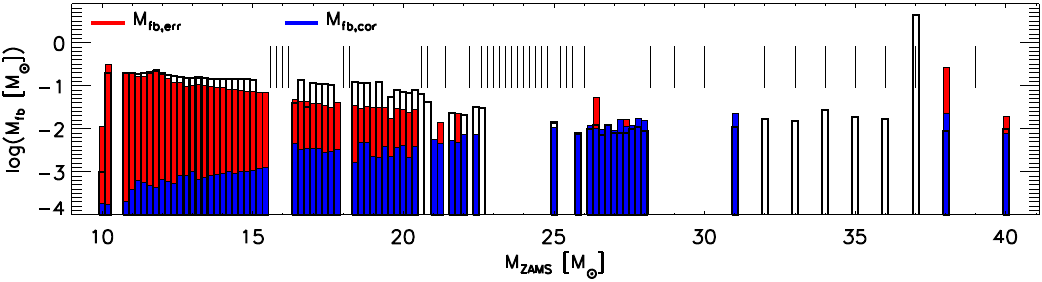}
\caption{Fallback masses for all progenitors investigated by
\citet{Ugliano2012}. The white histogram bars display the results of
\citet{Ugliano2012}. The red bars show the results from simulations by
\citet{Ertl2015} with modeling improvements in various aspects (but with
the same 19.8\,$M_\odot$ red supergiant model for the SN~1987A calibration
case), also applying the incorrect fallback estimate of Eq.~(\ref{eq:error}).
The blue histogram bars display the fallback masses as computed by
\citet{Ertl2015} with the correct fallback determination according to
Eq.~(\ref{eq:corrected}).
The vertical lines in the upper part of the plot indicate non-exploding 
cases obtained with the improved modeling by \citet{Ertl2015}.
Note that for better comparison with \citet{Ugliano2012} the models
reported in this plot do {\em not} include the calibration of the 
low-mass explosions with the Crab-like progenitor introduced by
\citet{Ertl2015} and \citet{Sukhbold2015}.
}
\label{fig:m_fallback_bar}
\end{center}
\end{figure*}
%------------------------------------------------------------------------

%-----------------------------------------------------------------------
\begin{figure*}
\includegraphics[width=\textwidth]{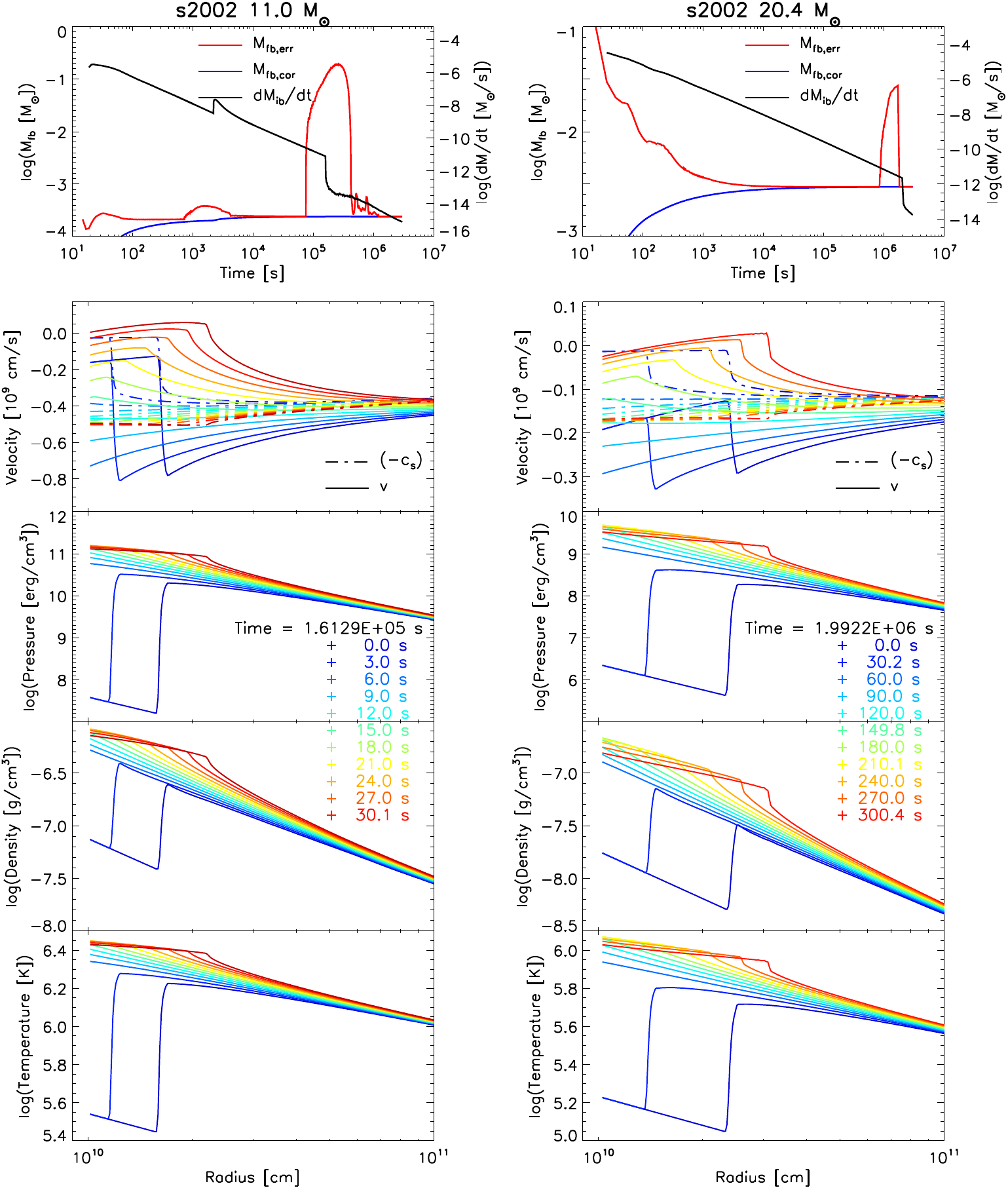}
\caption{Fallback and dynamical evolution in two exemplary explosion
simulations for an 11.0\,$M_\odot$ progenitor ({\em left}) and a
20.4\,$M_\odot$ case ({\em right}). The {\em upper panels} display
the total fallback mass as function of time. 
The red line gives the incorrect result from
applying the rhs of Eq.~(\ref{eq:error}) at all times.
The ``reflection time'', $t_\mathrm{r}$, is at about the instant of
the late, local maximum of the red curve. The previous, incorrect
estimate of the fallback mass (Eq.~\ref{eq:error})
corresponds to a value close to this maximum. The rising blue line
shows the correct evaluation according to Eq.~(\ref{eq:corrected}),
the black line with the decaying trend the corresponding 
mass-accretion rate, $\dot M_\mathrm{ib}$. The temporary increase 
of $\dot M_\mathrm{ib}$ in the 11.0\,$M_\odot$ model 
at $\sim$2000\,s is connected to the 
inward acceleration of matter by the first reverse shock that
is created when the supernova shock passes the carbon-helium 
interface. 
The sets of {\em lower panels} display velocity ($v$; 
solid lines) and adiabatic sound speed ($c_\mathrm{s}$; dash-dotted 
lines), pressure, density, and temperature profiles ({\em from top to
bottom}) for selected instants around the ``reflection time''.
}
\label{fig:m_fallback_examples}
\end{figure*}
%------------------------------------------------------------------------

An erroneous interpretation of the hydrodynamical results
led to an incorrect determination of the fallback masses in
\citet{Ugliano2012}, which also (on a smaller level) affects
the neutron star masses provided in that paper.
This problem was already addressed and corrected in the
follow-up works by \citet{Ertl2015} and \citet{Sukhbold2015}.
Therefore, the reader is advised to use the new data of the
latter two publications. In the remaining text of this 
{\em Erratum} we present the differences of the old and new 
fallback results in detail and explain the origin of the
mistake in the original analysis by \citet{Ugliano2012}.

The fallback masses obtained by \citet{Ugliano2012} are
displayed in Fig.~\ref{fig:m_fallback_bar} by white 
histogram bars. The red bars show the fallback masses
determined for the same progenitor set with the improved
explosion modeling of \citet{Ertl2015} and the same
{\em incorrect} fallback criterion as applied by \citet{Ugliano2012}.
The results basically agree, and a trendency of higher fallback
masses for lower-mass progenitors is present in both 
data sets. The quantitative differences are a consequence
of improvements of some modeling aspects by \citet{Ertl2015}.
The blue histogram bars represent the fallback masses as
determined with a {\em correct} evaluation of the fallback
for the models of \citet{Ertl2015}.
The correct values are considerably smaller in particular for 
progenitors below $\sim$20\,$M_\odot$. The trend of increasing
fallback masses for less massive stars is inverted to the
opposite behavior.

The reason for the error in the fallback analysis by
\citet{Ugliano2012} can be understood from the dynamics
plotted in Fig.~\ref{fig:m_fallback_examples}.
After the subsiding of the neutrino-driven wind (on a time scale
of 10--20 seconds after bounce), gas in the inner regions of the 
exploding star is decelerated by the gravitational pull of the
neutron star and collapses back. This leads to fallback with a
rate that peaks at some ten seconds post bounce and decreases
afterwards according to a power law ($\propto t^{-5/3}$).

On a time scale of days to weeks, the reverse shock created
when the supernova shock passes the helium-hydrogen interface,
propagates toward the center of the exploding star.
Moving backward through the inner layers of the star, 
the reverse shock accelerates the inward motion of the gas.
At the same time it compresses and heats the shocked gas, 
raising its pressure by 2--3 orders of magnitude. 
When the reverse shock leaves the computational domain
through the inner grid boundary, which is treated as an
open (outflow) boundary with a radial location of typically
$10^{10}$\,cm during these late stages of the evolution,
the gas in a large volume of the star has negative (supersonic)
velocities (Fig.~\ref{fig:m_fallback_examples}). Shortly
afterwards, however, the inflow of the stellar matter is 
reversed and a strong outward moving shock develops.

\citet{Ugliano2012} calculated the fallback mass by taking
at this ``reflection time'', $t_\mathrm{r}$,
the sum of all gas mass that had fallen 
through the inner grid boundary until this time plus an estimate
of additional fallback that will be added later from the mass
that has been accelerated inward by the reverse shock. For this
latter contribution they took the arithmetic average of the
mass at time $t_\mathrm{r}$ with negative velocities and the
mass with velocities smaller than the escape velocity
$v_\mathrm{esc}$ (in all cases these two masses were nearly
identical):
\begin{equation}
M_\mathrm{fb,err} = \int_0^{t_\mathrm{r}}\mathrm{d}t\,
\dot{M}_\mathrm{ib}(t) 
+ 0.5\,\left[M_{v < 0}(t_\mathrm{r}) + 
M_{v < v_\mathrm{esc}}(t_\mathrm{r})\right] \,.
\label{eq:error}
\end{equation}
The values of the fallback mass thus obtained are near the
local, late-time maxima of the red lines in the upper two
panels of Fig.~\ref{fig:m_fallback_examples}. 
Applying this recipe, \citet{Ugliano2012} erroneously assumed
that the dynamical evolution as computed for $t > t_\mathrm{r}$
is not trustworthy, because they interpreted the outward going wave 
as a consequence of a reflection of the reverse shock at the inner
grid boundary and therefore as a numerical artifact caused by the
presence of this boundary.

A detailed, time-dependent analysis, however, 
reveals that this interpretation
was not correct. Figure~\ref{fig:m_fallback_examples} shows
what happens. The infall of the reverse-shock heated matter is
decelerated because of the steepening of the negative 
pressure gradient
that happens as the inward flow gets geometrically focussed.
The deceleration produces a wave that begins to move outward
again and steepens into a shock front when the expansion velocity
exceeds the local sound speed. This happens well after the 
reverse shock has left the grid, at which time no numerical 
artifact is created (in fact, such an artifact would not be
able to travel into the computed volume because the postshock flow 
streams to the boundary with supersonic speed). 
The ``reflection wave'' is
therefore {\em not} a numerical artifact but a physics phenomenon,
in which the contracting stellar mass itself reverses its infall.
Tests with smaller radii for the location of the 
inner grid boundary (e.g., $5\times 10^8$\,cm or $10^9$\,cm) 
show exactly the same
dynamical behavior, confirming this ``self-reflection''.
The correct evaluation of the fallback is therefore a simple
time integral of the mass flow rate through the inner boundary:
\begin{equation}
M_\mathrm{fb,cor} = 
\int_0^\infty\mathrm{d}t\,\dot{M}_\mathrm{ib}(t)\,.
\label{eq:corrected}
\end{equation}
This integration was applied by \citet{Ertl2015} and 
\citet{Sukhbold2015} and yields the values of the blue
histogram bars in Fig.~\ref{fig:m_fallback_bar}.

\bibliographystyle{apj}

\begin{thebibliography}{}

\bibitem[Ertl et al.(2015)]{Ertl2015}
Ertl, T., Janka, H.-Th., Woosley, S. E., Sukhbold, T., and Ugliano,
M. 2015, \apj, in press; arXiv:1503.07522

\bibitem[Sukhbold et al.(2015)]{Sukhbold2015}
Sukhbold, T., Ertl, T., Woosley, S.~E., Brown, J.~M., \& Janka, H.-T.\
2015, \apj, submitted; arXiv:1510.04643

\bibitem[Ugliano et al.(2012)]{Ugliano2012}
 Ugliano, M., Janka, H.-T., Marek, A., \& Arcones, A.\ 2012, \apj, 757, 69

\end{thebibliography}

\end{document}